\documentclass{aipproc}
\layoutstyle{6x9}
\usepackage{bm}
\begin{document}

\title{Baryon spectroscopy on the lattice: recent results}

\author{Colin Morningstar}{
  address={Department of Physics, Carnegie Mellon University, Pittsburgh,
           PA, USA  15213-3890}}

\begin{abstract}
Progress in determining the baryon spectrum using computer simulations of
quarks and gluons in lattice QCD are summarized  and some future plans are
outlined.
\end{abstract}

\maketitle

Baryon spectroscopy is plagued by numerous unresolved issues.  The
quark model predicts many more states\cite{isgurkarl,capstick} than are
currently known.  Experiments in Hall B at Jefferson
Laboratory are currently mapping out of the spectrum of $N^\ast$ nucleon
excitations, so the question of the so-called ``missing resonances'' should
soon be resolved. A quark-diquark picture of baryons predicts a
sparser spectrum\cite{diquark}.   Various bag and soliton models have
also attempted to explain the baryon masses.  The natures of the Roper
resonance and the $\Lambda(1405)$ remain controversial.  Experiment
shows that the first excited positive-parity spin-$1/2$ baryon lies below
the lowest-lying negative-parity spin-$1/2$ resonance, a fact which is
difficult to reconcile in quark models.  First principles studies of the
baryon spectrum using lattice Monte Carlo methods are long overdue.

\begin{figure}
\begin{minipage}[b]{65mm}
\includegraphics[width=62mm,bb=31 277 524 683]{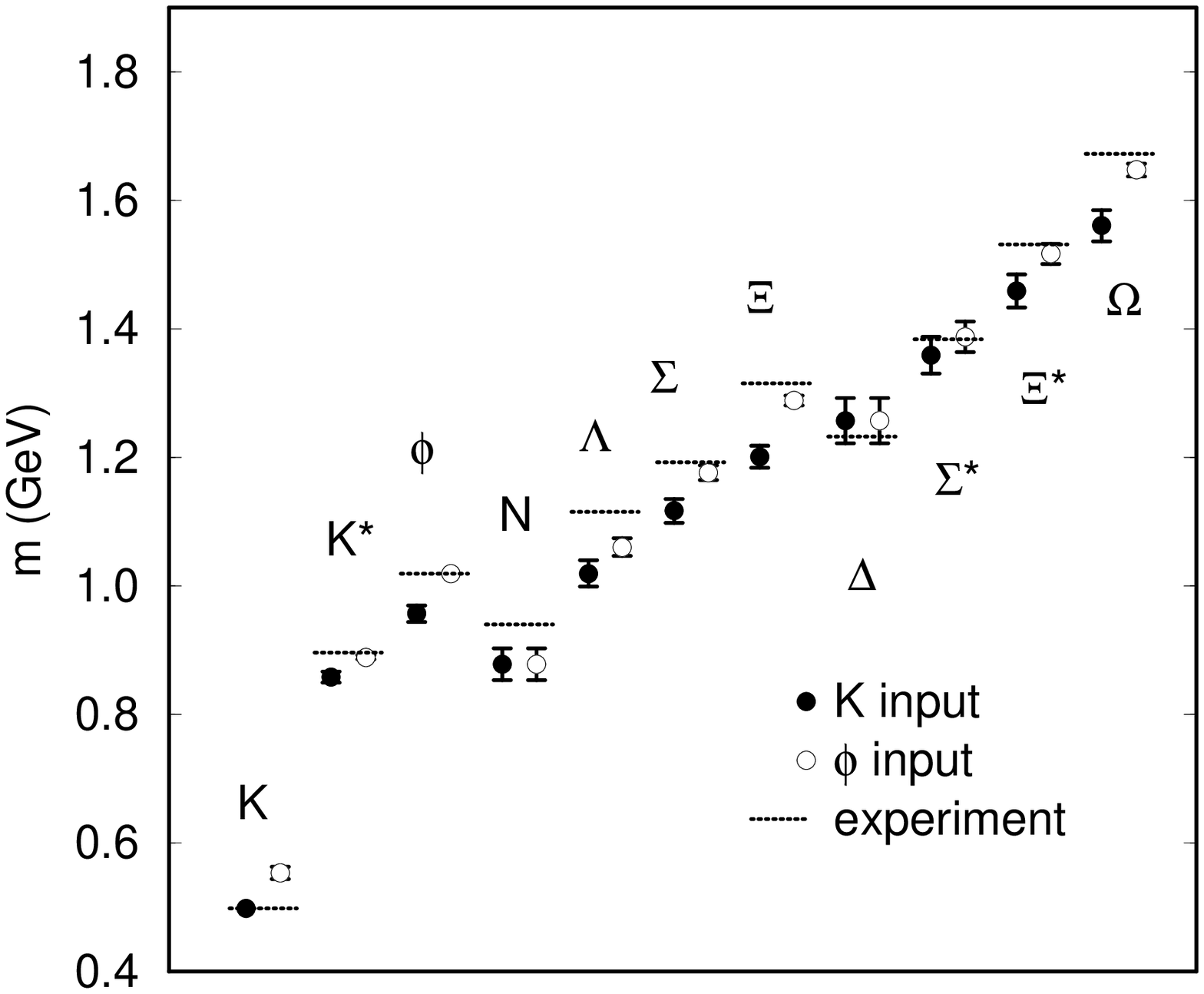}
\includegraphics[width=62mm,bb=38 80 536 470]{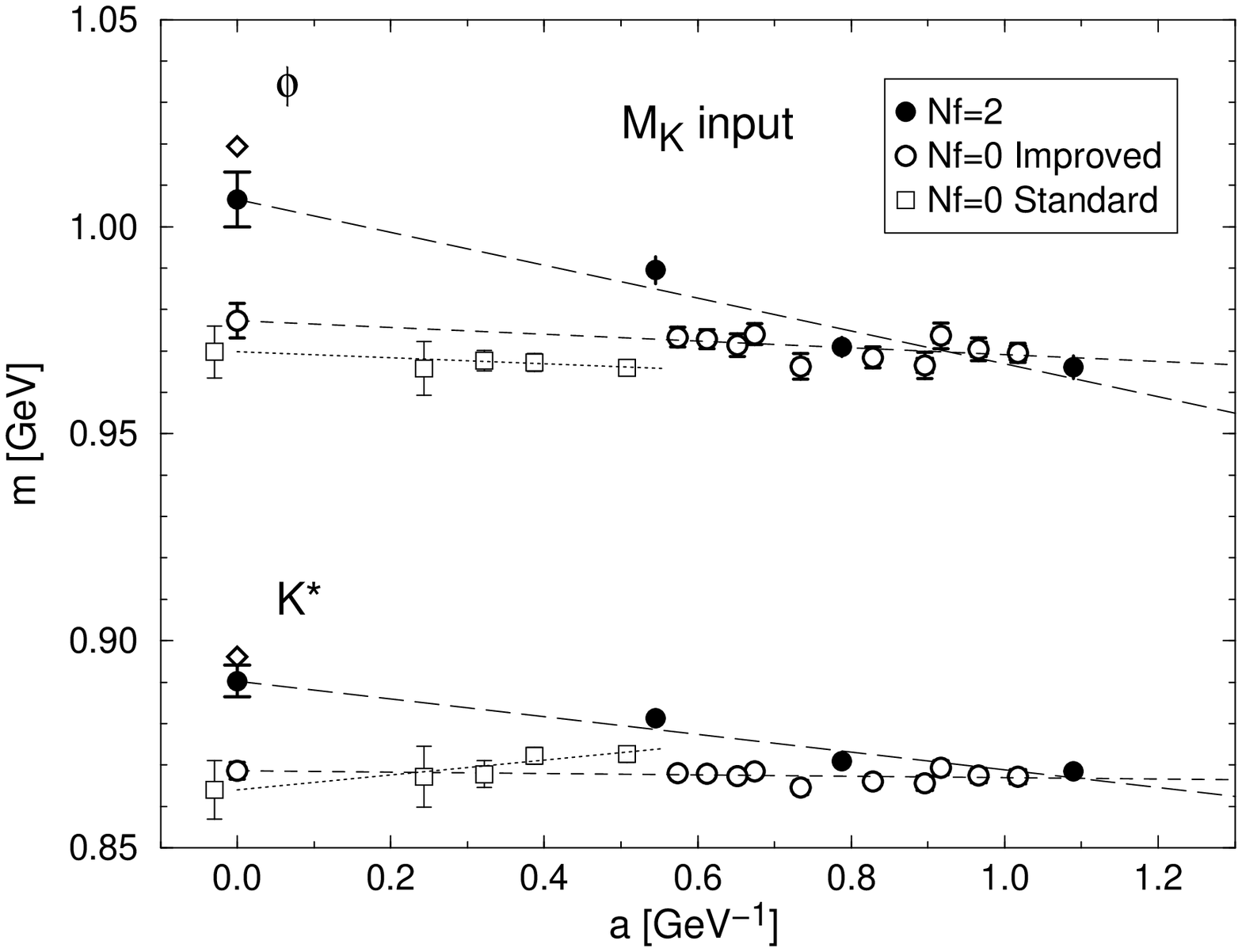}
\end{minipage}
\raisebox{6mm}{\includegraphics[width=76mm,bb=48 82 507 646]{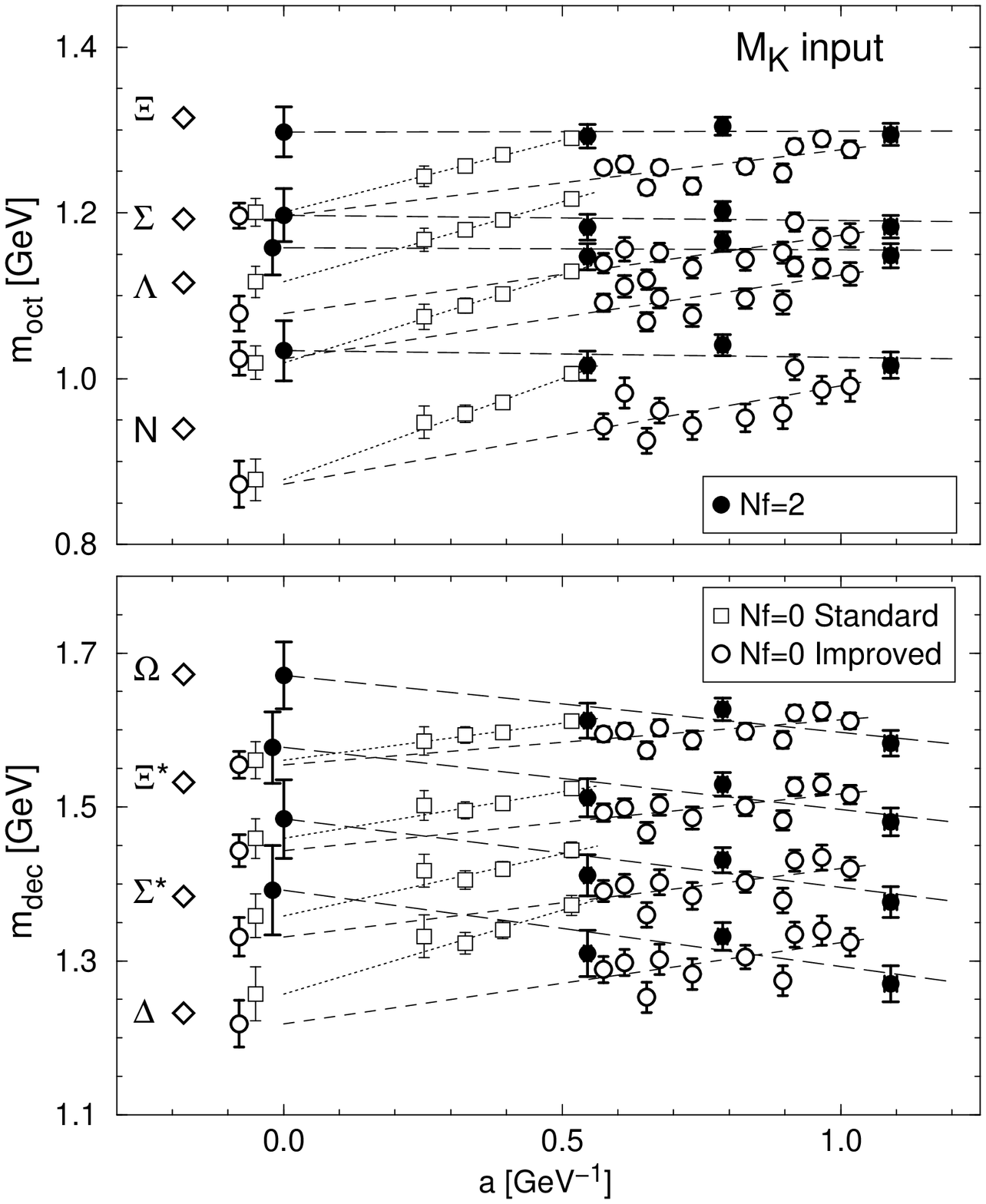}}
\caption{ State of the art results from the CP-PACS collaboration
 for the low-lying hadron spectrum.
 Quenched results (upper left) in the continuum limit are from 
 Ref.~\protect\cite{CPPACS1}.  Solid symbols use the $K$ to set the scale,
 and the hollow symbols use the $\phi$ to set the scale.  The differences
 between the two sets measure the systematic errors from quenching.
 Agreement with experiment is remarkable, indicating that quenching errors
 in these observables are not large.  In the lower right and the
 two leftmost plots, solid symbols indicate results from 
 Ref.~\protect\cite{CPPACS2} including two flavors
 of light quark loops, whereas open symbols are quenched results shown
 against the lattice spacing $a$.  For the $K$ and $\phi$ mesons, one
 observes excellent agreement with experiment.  However, the baryon
 results are problematic; the authors suggest that finite volume effects
 are to blame.  Experimental results are indicated by the diamonds.
\label{fig:CPPACS}}
\end{figure}

State of the art results for the low-lying hadron spectrum using the
Iwasaki gauge action and a clover tadpole-improved fermion action are
presented in Fig.~\ref{fig:CPPACS}.  The quenched spectrum (upper left)
deviates from experiment by under ten per cent.  The inclusion of
two flavors of light quark loops produces excellent agreement with
experiment for the $K$ and $\phi$ mesons, but the baryon masses
show significant deviations from their experimental values.  The authors
suggest that finite volume errors will explain these discrepancies.

During the last few years, a handful of lattice studies
have begun at last to focus attention on the excited baryon spectrum.
Other than some preliminary unquenched results, all estimates to date
have utilized the quenched approximation, 
and most have used unphysically heavy quarks.
Systematic errors due to discretization and finite volume are not yet under
control.  Although the current status of such calculations is still embryonic,
a renewed interest of the lattice community
in such calculations promises substantial progress in the near future.
For example, the Lattice Hadron Physics Collaboration (LHPC)\cite{lhpc}
recently formed and one of its major objectives is the computation of the
$N^\ast$ spectrum.  The formation of this collaboration was spearheaded
by the late Nathan Isgur and is funded by the Department of Energy's
Scientific Discovery through Advanced Computing (SciDAC) initiative.
I shall outline the plans of this collaboration later in this talk.
But first, results from four selected recent baryon studies are presented.

\begin{figure}
\begin{minipage}[t]{142mm}
\includegraphics[height=70mm,angle=90,bb=51   64  528  735]{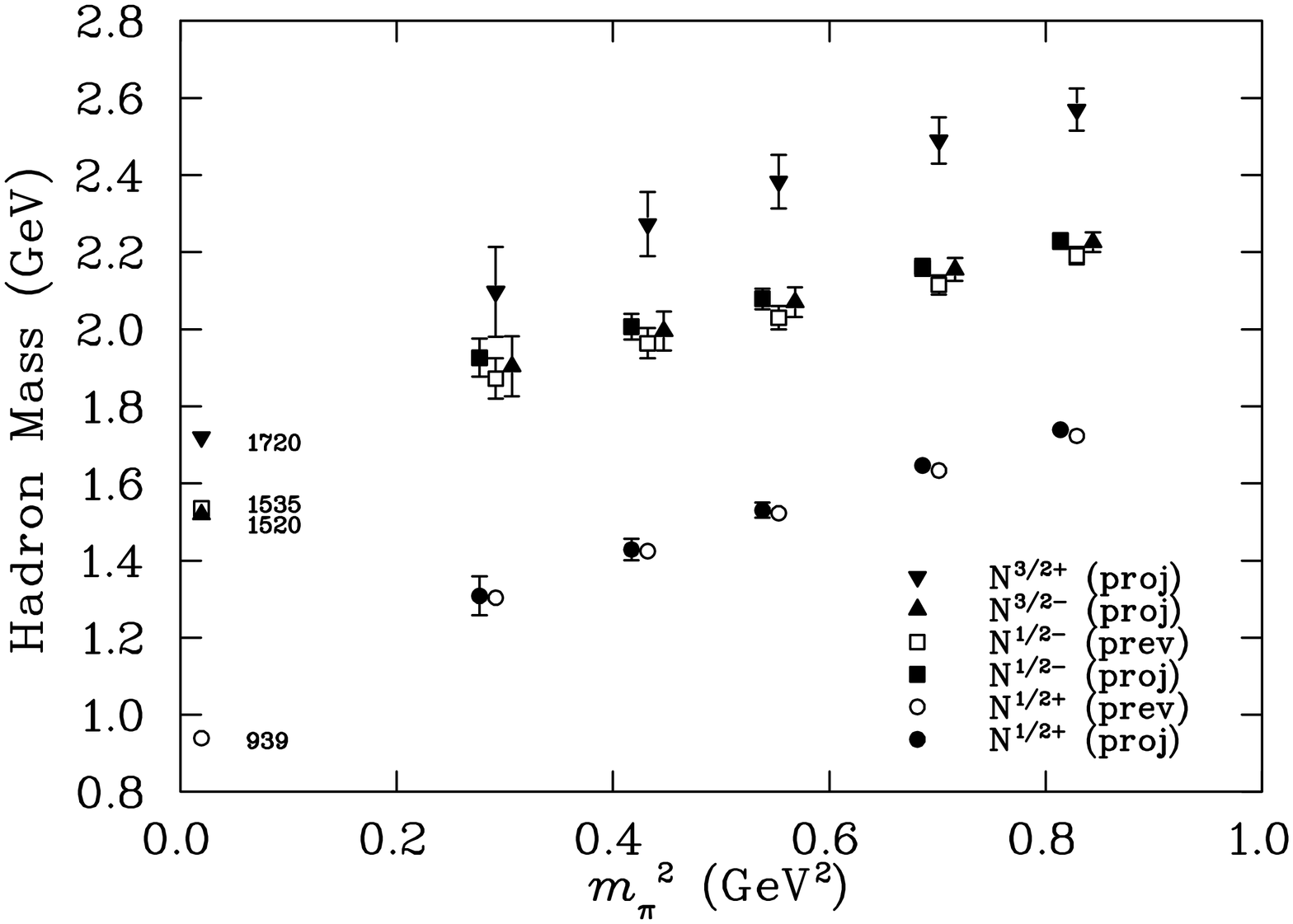}
\includegraphics[height=70mm,angle=90,bb=51   64  528  735]{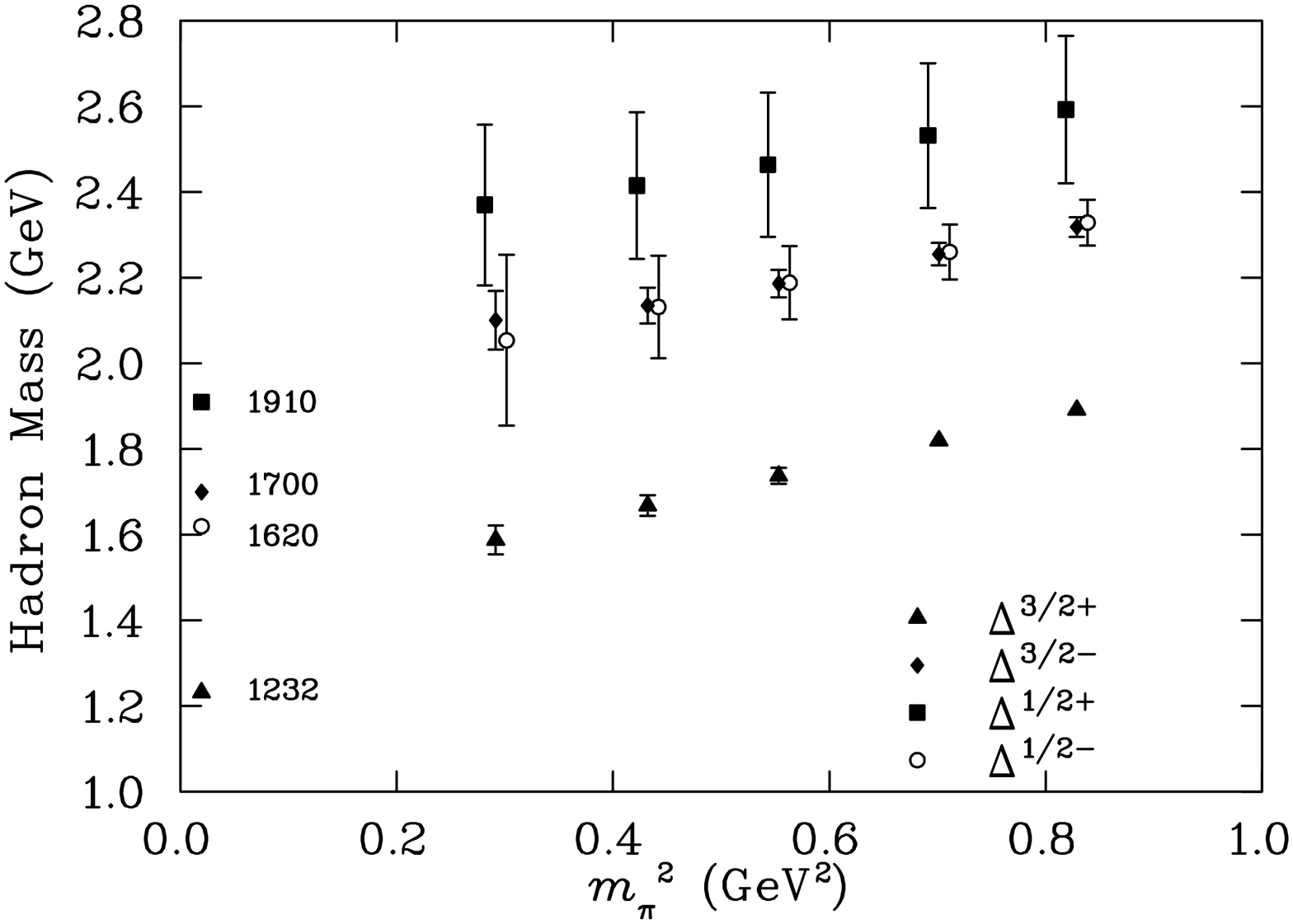}
\end{minipage}
\caption{Mass estimates of the $J^P=\frac{1}{2}^\pm$ and
 $\frac{3}{2}^\pm$ $N$ and $\Delta$ baryons from Ref.~\protect\cite{adelaide}
 against the square of the pion mass $m_\pi$ in the quenched approximation.
 The results use an improved gauge field action and the fat-link irrelevant
 clover (FLIC) fermion action on a $16^3\times 32$ lattice with spacing
 $a=0.12$ fm, set by the string tension. Spin-projected results are compared
 with previous unprojected ones. Experimental values are shown near
 the vertical axis.
\label{fig:adelaide}}
\end{figure}

Latest results in the quenched approximation from the CSSM Lattice collaboration
using an improved gauge field action and a fat-link irrelevant clover (FLIC)
fermion action are shown in Fig.~\ref{fig:adelaide}. Rather heavy quark masses
were used, but the level orderings are in qualitative agreement with
those observed in experiments.

The first-excited state in the positive-parity spin-$1/2$ sector
is found to be significantly higher\cite{sasaki0,dgr,glozman,adelaide0}
than the Roper mass when unphysically large quark masses are used.
\begin{figure}
\begin{minipage}[t]{125mm}
\includegraphics[width=62mm,bb=74  126  538  570]{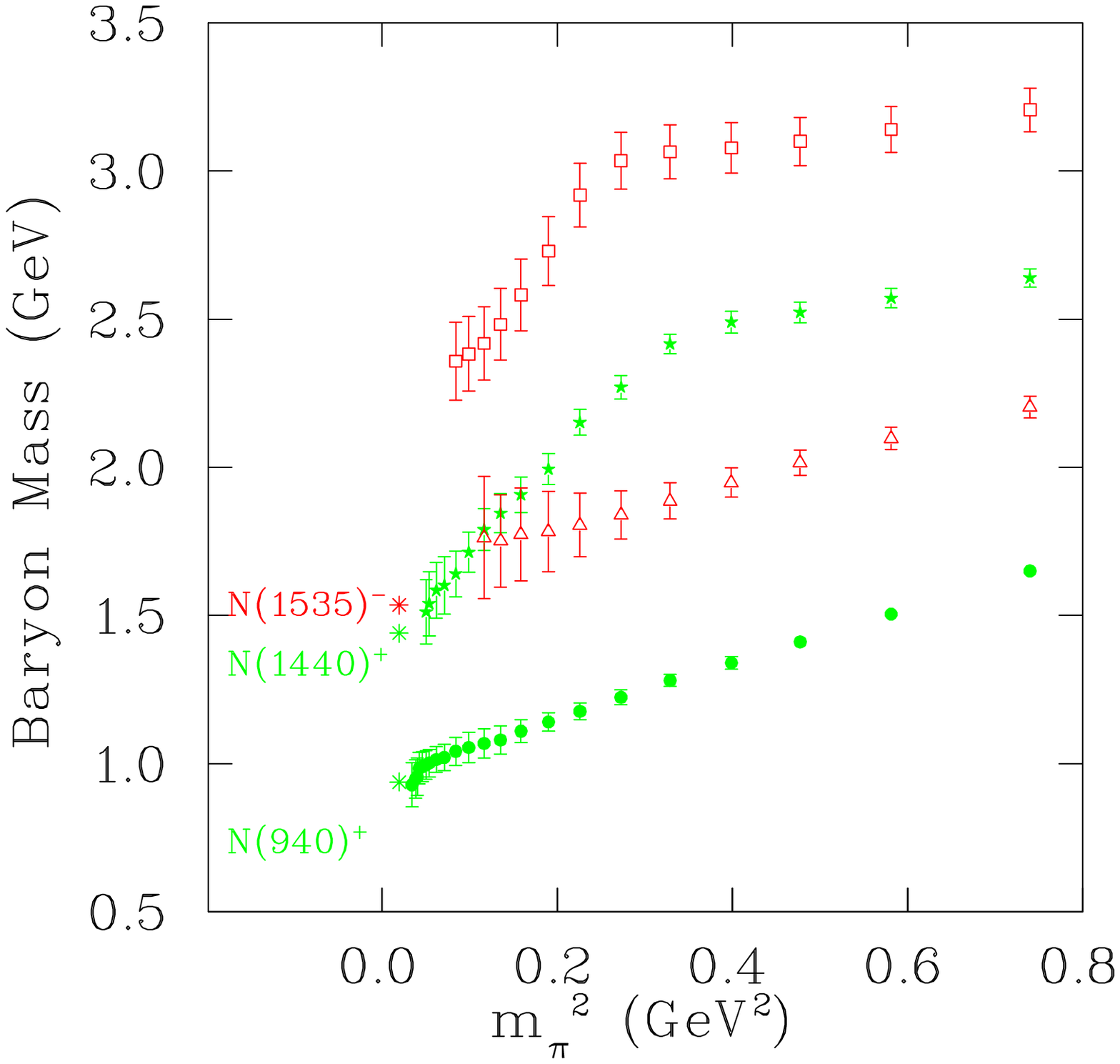}
\includegraphics[width=62mm,bb=74  126  538  570]{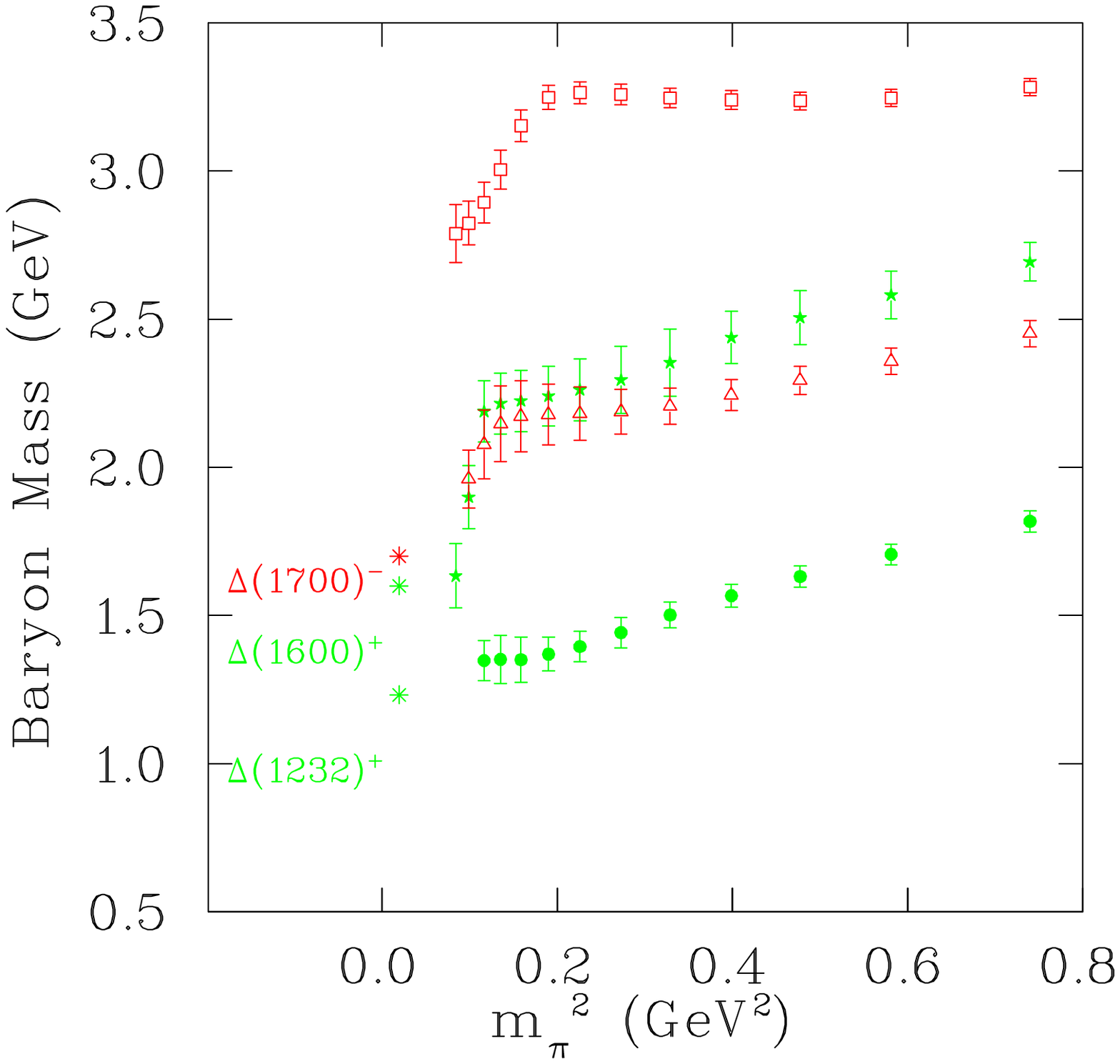}
\includegraphics[width=62mm,bb=74  126  538  570]{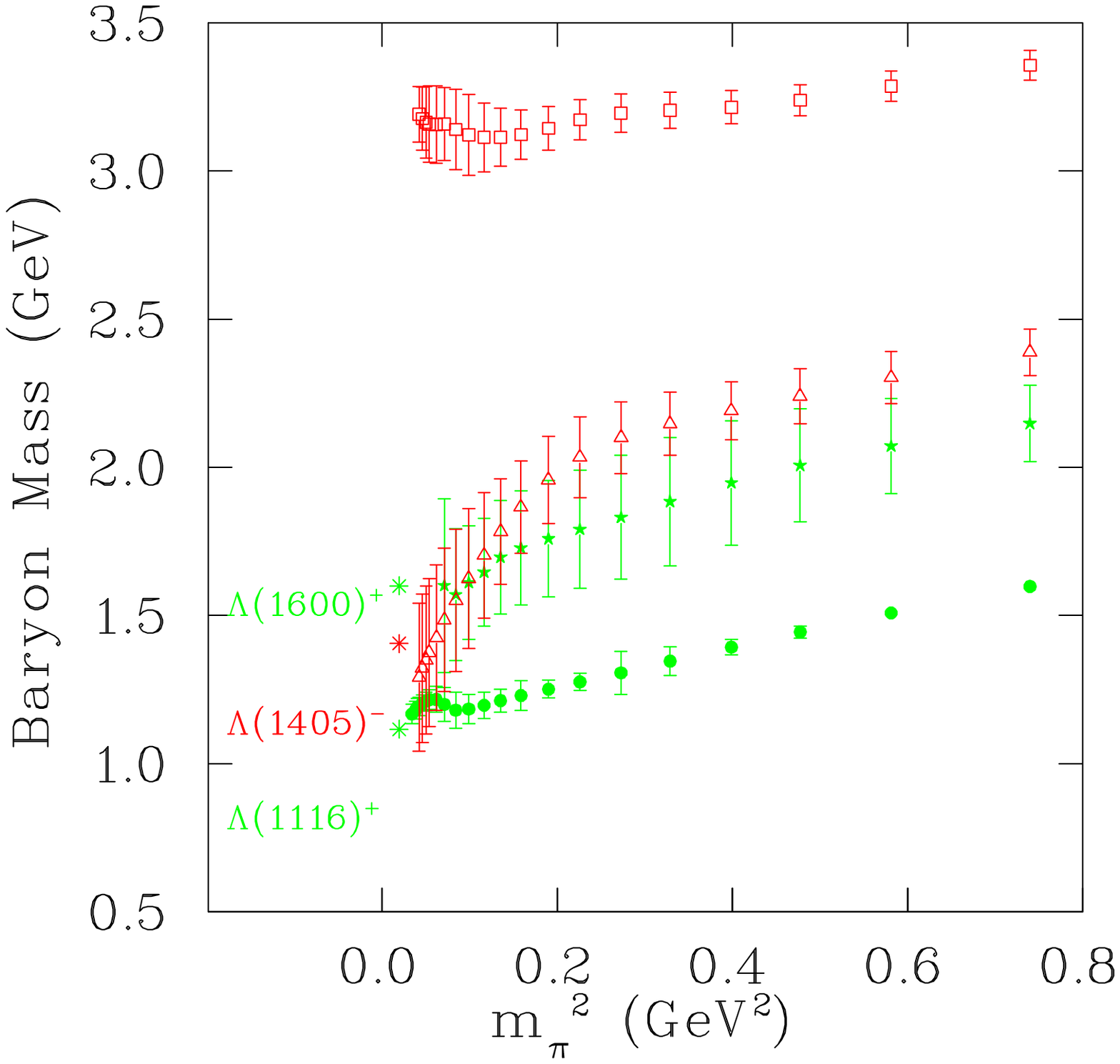}
\includegraphics[width=62mm,bb=74  126  538  566]{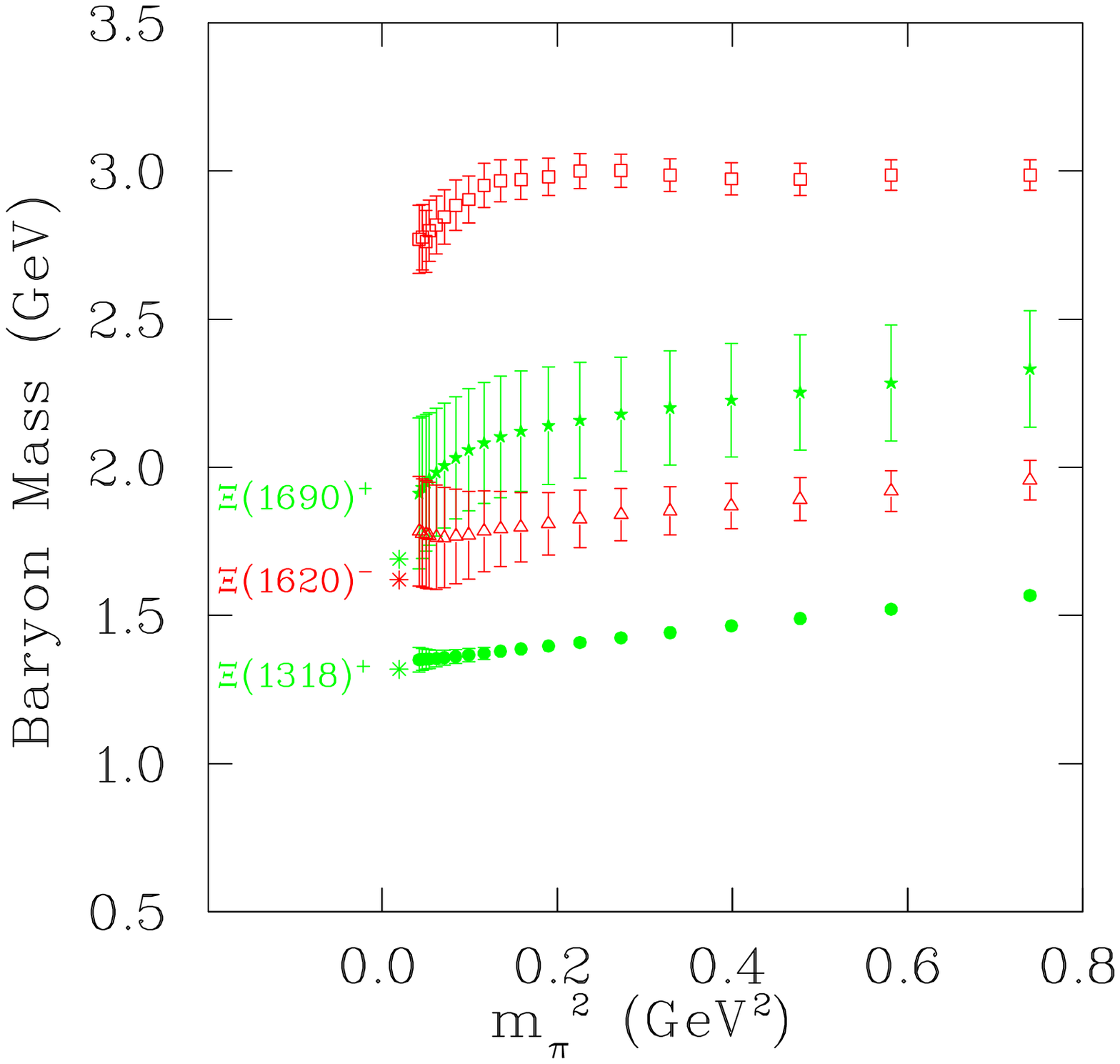}
\end{minipage}
\caption{ Two lowest-lying octet and decuplet baryon masses for both
 positive and negative parities in the quenched approximation
 from Ref.~\protect\cite{lee} at lattice spacing $a\approx 0.2$ fm
 against the square of the pion mass.  Results were obtained on 
 $16^3\times 28$ lattices using an improved gauge action and overlap
 fermions for a large range of light quark masses.  These figures emphasize
 the importance of simulating with sufficiently light quark masses.
 Experimental measurements are shown as bursts.
\label{fig:lee}}
\end{figure}
Recently, the use of overlap fermions has allowed quenched
calculations\cite{lee} with
realistically light quark masses, and this point appears to be crucial
for identifying the Roper as a radial excitation of the nucleon. 
Fig.~\ref{fig:lee} shows dramatic changes in the quenched
baryon masses as the pion mass drops
below 300 MeV.  An important note of caution concerning these findings is
the use of empirical Bayesian constrained curve fitting to extract the
excited state mass from a single correlation function.  Although
likely reliable, an analysis using several operators in a correlation matrix
would be much preferred.  Also, at such light quark masses, one must
very carefully check finite volume errors (as evidenced in the next study
described below). In a more recent paper\cite{lee2}, these authors have also
addressed the issue of pollution of the first-excited state observed in
Fig.~\ref{fig:lee} by unphysical $\eta^\prime N$ quenched artifacts.  Such 
ghost contributions were distinguished by obtaining results in two
volumes $16^3\times 28$ and $12^3\times 28$, corresponding to lattice extents
$3.2$ and $2.4$ fm, respectively.  The conclusion, within the quenched
approximation, that the Roper is a
radial excitation of the nucleon with three valence quarks was confirmed.

\begin{figure}
\begin{minipage}[t]{142mm}
\includegraphics[width=70mm,bb=67 67 597 566]{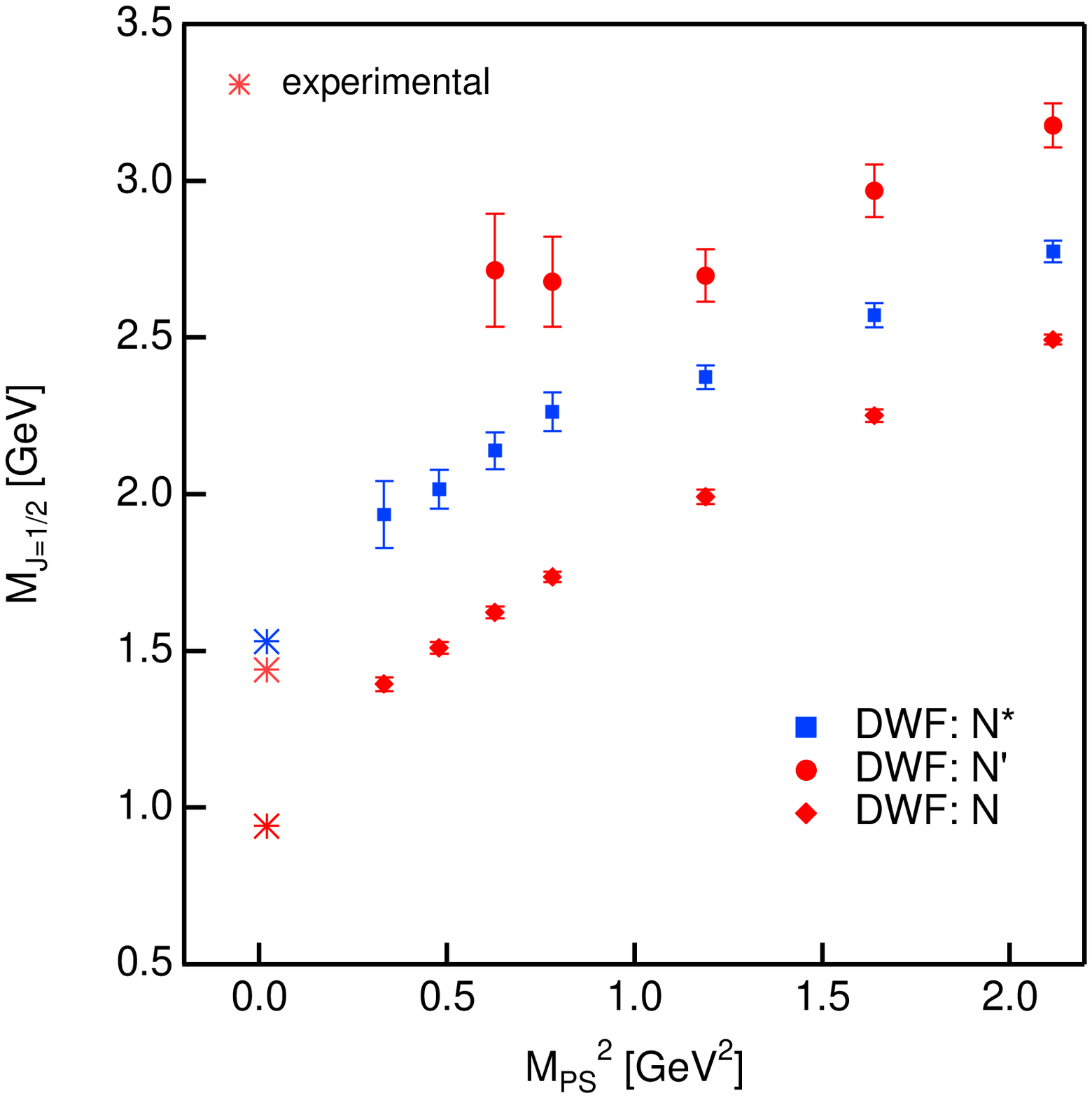}
\includegraphics[width=70mm,bb=67 67 597 566]{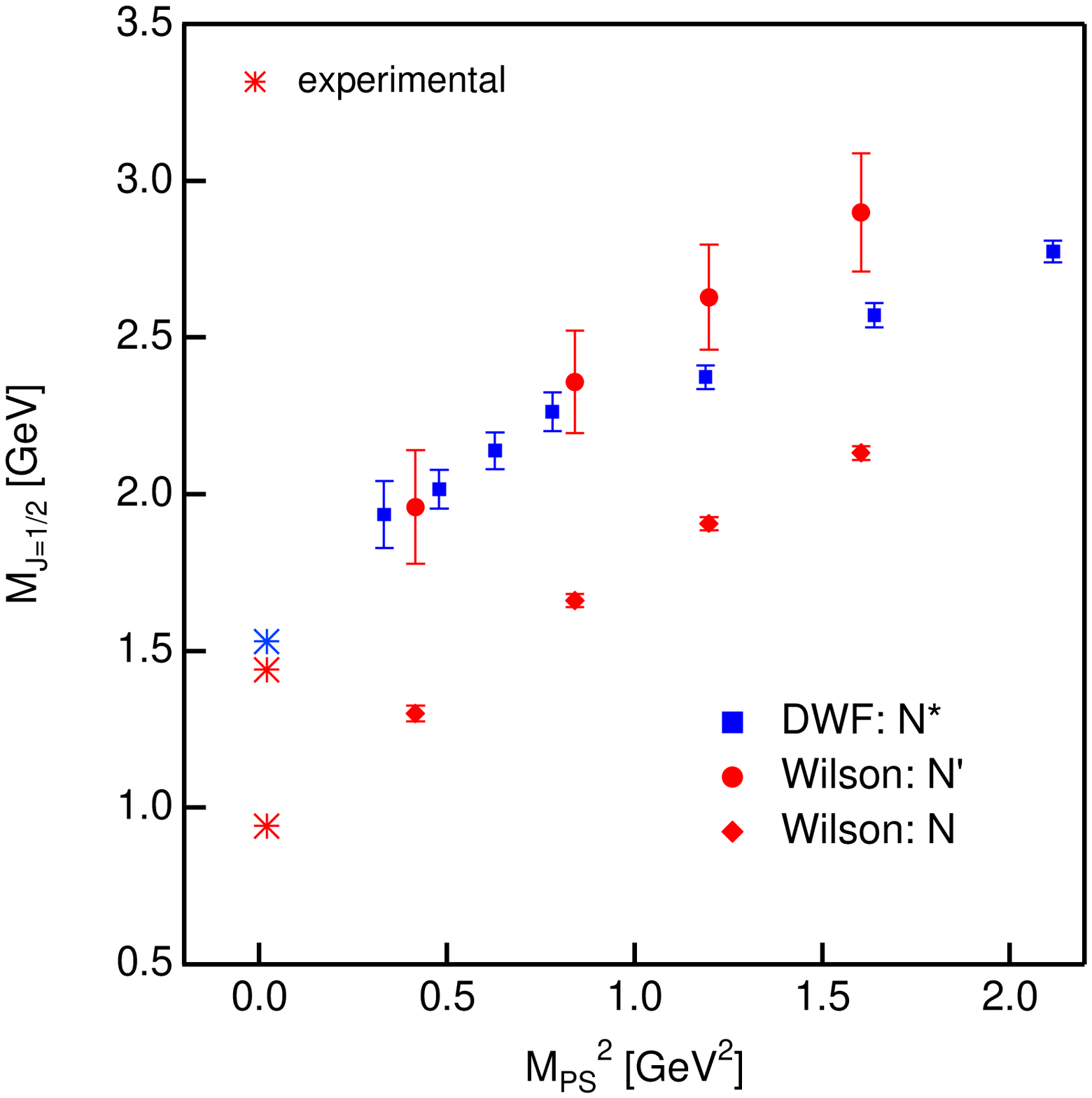}
\end{minipage}
\caption{The low-lying nucleon masses in the quenched approximation
 from Ref.~\protect\cite{sasaki0} (left) and Ref.~\protect\cite{sasaki}
 (right).  Results on the left use domain wall fermions in a small
 volume with spatial extent $La\approx 1.5$ fm. On the right,
 results in three different volumes $La\approx 2.2 - 3.0$ fm
 were extrapolated using $1/L^3$ to the infinite volume limit.
 The Wilson gauge and Wilson fermion actions with $\beta=6.0$ and spacing
 $a\sim0.1$ fm were used with quark masses yielding pion masses in
 the range $m_\pi=0.6-1.0$ GeV.  Maximum entropy methods were employed.
 Experimental values are shown as bursts.
\label{fig:sasaki}}
\end{figure}

A large sensitivity of the Roper resonance to finite volume errors
in the quenched approximation has recently been reported in 
Ref.~\cite{sasaki}.  In Fig.~\ref{fig:sasaki}, one sees that 
for $m_\pi^2\approx 0.5$ GeV$^2$, the infinite volume results for the
$N^\prime$ Roper are degenerate with the negative parity $N^\ast$, in
disagreement with the results shown in Fig.~\ref{fig:lee}.  The lattice spacings
and actions differ, so this discrepancy could simply be a discretization
artifact.  Also, maximum entropy methods are being employed, which further
muddies the issue.  Nevertheless, an important message seems clear: large
volumes and small quark masses are especially important for reliable results
in baryon spectroscopy.

Doubly charmed baryons, also of current experimental interest, have
also been studied recently\cite{lewis} in the quenched approximation
using an improved gauge action on anisotropic lattices with the D234 action
for the light quarks and a nonrelativistic (NRQCD) action for the heavy
quarks.  Two lattice spacings $a\sim 0.15, 0.22$ fm and four light quark
masses were used.  These authors found that
mass splittings between $J=\frac{1}{2}$ and $\frac{3}{2}$ baryons from color
hyperfine interactions were {\em not} suppressed, unlike the meson sector.
Many charmed and bottom baryons were studied.
No finite volume checks were done, and radiative corrections to the
couplings in the NRQCD action were ignored.  Results using a clover
fermion action at $\beta=6.2$ have also been presented\cite{flynn} recently.

The Lattice Hadron Physics Collaboration is currently using
large sets of extended operators with correlation matrix techniques
to capture a significant portion of the baryon spectrum.  Excited states
will be extracted without resorting to maximum entropy methods.
We believe that the construction of good operators is crucial: the operators
have been designed with one eye towards maximizing overlaps with the low-lying
states of interest, and the other eye towards minimizing the number of
source needed in computing the required quark propagators.  For example,
the three-quark operators we plan to use are expressed in terms of
smeared quark fields $\tilde{\psi}$, the covariant three-dimensional
Laplacian $\tilde{\Delta}$, and the $p$-link covariant displacement
$\tilde{D}^{(p)}_j$ by
\begin{center}
\begin{tabular}{ll}
(1): & $\!\!\!\!\!\!\phi^F_{ABC} \varepsilon_{abc} \Gamma_{\alpha\beta\gamma}
  \!\bigl( \tilde{\Delta}^{n_1} \tilde{\psi}\bigr)_{Aa\alpha} 
  \!\bigl( \tilde{\Delta}^{n_2} \tilde{\psi}\bigr)_{Bb\beta}
  \!\bigl( \tilde{\Delta}^{n_3} \tilde{\psi}\bigr)_{Cc\gamma}$, \\
(2): & $\!\!\!\!\!\!\phi^F_{ABC}\ \varepsilon_{abc}\ \Gamma_{\alpha\beta\gamma}^{j}
 \ \!\bigl( \tilde{\Delta}^{n_1} \tilde{\psi}\bigr)_{Aa\alpha}
 \ \!\bigl( \tilde{\Delta}^{n_2} \tilde{\psi}\bigr)_{Bb\beta}
 \ \!\bigl( \tilde{D}^{(p)}_j\!\tilde{\Delta}^{n_3} 
           \tilde{\psi}\bigr)_{Cc\gamma}$, \\
(3): & $\!\!\!\!\!\!\phi^F_{ABC}\ \varepsilon_{abc}\ \Gamma_{\alpha\beta\gamma}^{jk}
 \ \!\bigl( \tilde{\Delta}^{n_1} \tilde{\psi}\bigr)_{Aa\alpha} 
 \ \!\!\bigl( \tilde{D}^{(p_1)}_j\!\tilde{\Delta}^{n_2} 
           \tilde{\psi}\bigr)_{Bb\beta}
 \ \!\!\bigl( \tilde{D}^{(p_2)}_k\!\tilde{\Delta}^{n_3}
           \tilde{\psi}\bigr)_{Cc\gamma}$,
\end{tabular}
\end{center}
where $n_1$, $n_2$, $n_3$, $p$, $p_1$, $p_2$ are positive integers, 
$j,k=\pm 1,\pm 2, \pm 3$ are spatial directions, $\alpha, \beta, \gamma$
are Dirac spin indices, $A, B, C$ are quark flavors, and $a,b,c$ indicate colors.
Different powers of the spatial Laplacian $\tilde{\Delta}$ are utilized to
build up radial structure, and the displacement operator $\tilde{D}_j$ 
is used to incorporate orbital structure.  The group theoretical projections
necessary to obtain operators transforming irreducibly under the symmetries of
the lattice have been carried out using software written in Maple.  Degeneracy
patterns among the different irreducible representations of the cubic
group must be exploited to identify angular momentum $J$ eigenstates in
the continuum limit.  We hope to present our first results in the near
future.  Note that our operator construction approach can be easily
adapted for mesons, pentaquark systems, and so on.

Regardless of how the operators are constructed, extracting the baryon
spectrum in lattice simulations remains a challenge.  It is especially
important for baryons that the Monte Carlo calculations be done in large volumes
with realistically light quark masses and without the quenched approximation. 
Furthermore, all baryon studies must ultimately confront the thorny issue of
treating unstable resonances. The techniques for doing this are well
known\cite{dewitt}, but are untested in QCD.
However, computing speeds continue to increase
and large computer clusters dedicated to hadron physics are coming on-line.
Given the renewed interest in baryon spectroscopy,
substantial progress is inevitable.
This work was supported by NSF award PHY-0099450.
\vspace*{-2mm}

\bibliographystyle{aipprocl}
\bibliography{baryons}
\end{document}